# Democratizing Making: Scaffolding Participation Using e-Waste to Engage Under-resourced Communities in Technology Design

Dhaval Vyas

The University of Queensland, Australia, d.vyas@uq.edu.au

Awais Hameed Khan

The University of Queensland, Australia, awaishameed.khan@uq.edu.au

Anabelle Cooper

The University of Queensland, Australia, anabelle.cooper16@gmail.com

## ABSTRACT

Maker culture and DIY practices are central to democratizing the design of technology; enabling non-designers (future end-users) to actively participate in the design process. However, little is known about how individuals from under-resourced communities and low socioeconomic status (SES) backgrounds, can practically leverage maker practices to design technology, creating value for themselves or their communities. To investigate this, we collaborated with an e-waste recycling centre, involving 24 participants (staff and low-SES volunteers) in two participatory maker workshop activities. Participants were provided with a generative e-waste toolkit, through which they repurposed e-waste materials and developed novel technology prototypes that created value from their perspectives and agendas. Our findings unpack three factors that influenced their making: balancing personal and community needs; incorporating convenience and productivity; and re-thinking sustainability and connection; and discuss strategies for scaffolding participation and engagement of under-resourced communities in making using an e-waste generative toolkit to democratize technology design.

# 1 Introduction

Making practices involving technological materials are generally conceptualised as methods of fostering innovation and learning [45,60]. Community plays a central role in the proliferation of maker culture—both physical (i.e., maker spaces) and online (i.e., forums, social media groups etc.) maker communities, encourage collaboration, peer-learning, and resource-sharing within this creative and technical environment [56]. Contemporary maker culture has expanded beyond its grassroots of anti-consumerist underpinnings to encompass entrepreneurial [36,45,52,64], therapeutic [12,48], educational [44,59], art [66,77] and sociopolitical [75] praxis. Making practices are also closely aligned to participatory design—where 'making', alongside telling and enacting, is a fundamental characteristic of tools and techniques of participation [7,70]; enabling participants to create externalisations, share ideas, and foster participatory mindsets [69]. Like maker culture, participatory design too has also evolved beyond its origins of workplace democratisation (See e.g. [19,23,46]) and now expanded into application across various domains and contexts [3] as a vehicle for democratising innovation [5,10]. These parallels in maker culture and participatory design have resulted in an increase in work that explores the intersections of making and participatory, co-creative approaches to generate ideas and products between makers and end users (e.g. see [44,59]).

While the overall making agenda promises democratization of products and concepts [76], it can arguably exclude under-resourced communities, who lack material resources, have limited technical abilities, and frequently struggle due to inherent systemic challenges [59,78]. With this increasing focus on maker practices within under-resourced communities in DIY and Making activities [73,86,87]—little is known about: *How to best engage individuals from under-resourced communities in community-based participatory maker activities? What kind of material scaffolding is required to foster this participation? How can individuals from under-resourced and low SES backgrounds practically leverage maker practices to design technology, creating value for themselves or their communities?*

To investigate this, we conducted two of participatory maker workshops in collaboration with an e-waste recycling social enterprise (hereafter referred to as 'the centre'), that provides work experience to unemployed low SES volunteers who are part of the Australian 'work for the dole'[1] (WFTD) program. Situated within a predominantly low SES catchment of a metropolitan city, the centre engaged WFTD participants in disassembly of e-waste; and employed skilled staff who engaged in electronic repair-work and repurposed e-waste to create new products. The two maker workshops, inspired by design workshops, hackathons and the Inventor Day [80] formats, engaged a total of 24 participants (20 WFTD participants, 4 centre staff) over a one-week period; and employed a generative e-waste toolkit to scaffold participation. The aim of these workshops was to investigate: How can individuals from under-resourced and low SES backgrounds practically leverage maker practices to design technology, creating value for themselves or their communities? Through the workshops, participants generated design concepts and technological prototypes that embodied their values; catering to individual interests and motivations, as well as addressing challenges and opportunities to improve their environments—both the centre, and broader community. Our findings show that the maker workshops provided a meaningful platform for the WFTD participants to explore technological solutions to their day-to-day issues. We found that our participants' engagement with e-waste opened up their potential for improving their current situations, supporting their local community and expressing artistic capabilities.

Building on the work of salvage fabrication [16,17] and participation in maker culture [12,24,75,87], this paper makes two principal contributions: First, we present empirical insights pertaining to motivations and concerns of under-resourced community members through their making process of technological prototypes in a participatory maker workshop context. Our findings unpack three factors that influenced making: balancing personal and community needs; incorporating convenience and productivity and re-thinking sustainability and art.

Second, we present strategies for scaffolding participation for under-resourced communities using e-waste generative toolkits. This covers reflections on how e-waste as a material, within the socio-political context of the centre enabled creative exploration, material 'back-talk' and reflexivity, and ownership and agency for participants to engage within the making process.

---

[1] Work for the dole is a government-funded welfare program in Australia where unemployed individuals spend time in labor intensive organizations to avail their welfare payments. https://www.dewr.gov.au/work-dole

## 2 Related Work

### 2.1 Maker Events, Hackathons and Communities

Makerspaces are often resource abundant environments, equipped with technical equipment (e.g., CNC machines, routers, specialised tools, and electronics) for rapid prototyping which require technical prowess; however the proliferation of low-cost, easy-to-use materials such as Arduinos, does lower some of the barriers to entry, however does not address learning challenges [45]. It is also important to note that makerspaces, as shown in HCI literature, are predominantly male dominated (with exceptions e.g., [24,30]), and mostly have participants from tech savvy backgrounds [13,56]. Community-based approaches such as design workshops [28,29] and hackathons [18,20,21,57] have been touted to engage members of under-resources communities in collaborative communal dialogue.

In addition to makerspaces, another popular participatory structure, that engages a broad community of members in making activities are hackathons. The hackathon, and similarly structured events are growing in popularity because of their format enables bringing together a diverse set of publics together in a participatory setting to explore ideas and solutions for specific issues. While the initial conceptualizations of hackathons focused on software development and writing code, more recent HCI studies have focused on topics including social justice issues [18], mental health [4], feminist designs [31], neighbourhood civic technologies [80,81], among others. The effectiveness of hackathons and their expected outcomes have often been questioned[57,79]; however Irani [37] argues that "*the hackathon rehearses an entrepreneurial citizenship*" to foster social change; citing it as "*one emblematic site of social practice where techniques from information technology production become ways of remaking culture*"—ergo it is more than what is 'made' during a hackathon; the social value of participating in one has significant benefits. A typical hackathon structure and its instantiations follow intense activities with 'manufactured urgency' [37]; associated with making, producing, and evaluating technologies or parts of them by involving a large group of like-minded people over one or two days. Specific adaptations and changes are commonly seen where experts and non-experts are grouped in the same sessions. For example, Taylor et al. [80,81] developed maker events called *Inventor Days* in order to garner grassroot innovation within a neighbourhood community. These maker events enabled community members to work with local makers to explore ways through which civic technologies can be designed for public spaces. Their work highlights the relationships and skills that are built by the community members which can be useful for sustaining such efforts for long term. The skills transfer and relationship building that happens over hackathons are useful means to have long-term and sustainable efforts for conducting collaborative efforts. Irani [37] applying a Latourian lens to hackathons, contends that they are "*a moment of design*", in that it is temporally situated, where urgency is created towards certain matters and cultural value triumphs over the tangible output. '*Maker Faire'* [50] are exhibition-like spaces where makers and DIY enthusiasms come together emphasizing their technological, social and economic interests while leaving space for practices that may be quite nuanced. The playful and explorative side of hackathons is also well-studied. Robinson and Johnson [63] showed how a hackathon helped put open data into public use and provided the local government staff with valuable feedback on application possibilities. We hence use a hackathon-inspired structure, that builds on the *Inventor Days* [80,81] format to structure and scaffold participation in the workshops in this study.

### 2.2 e-Waste, Repair and Design

The generative toolkit within this study uses e-waste materials to scaffold participation—we detail in this section how e-waste, repair and design intersect in contemporary maker and design work. Within HCI practice, working with waste has been shown to enable "*emerging forms of technology production via tools that position marginal, displaced, and discarded materials as central and useful again*" [17]. The creative uses of e-waste have been well studied in HCI [43], as well as how the disassembly of e-waste material takes place [53]. Serval studies have shown how upcycling can help individuals preserve memories and culture through creative reuse of waste or unwanted materials in their homes [85,89]. In fact, movements such as 'technological disobedience' [66], recyclism [27], and steampunk [77] have blurred the boundaries between repair, sustainability and art. The movement of technological disobedience was originated in Cuba where, due to its isolation, people started to challenge technological complexity and its exclusionary features. In this movement, people in Cuba created a new market of re-invented products and technologies that were created using craft approaches in a country where the idea of market was illusive. Rognoli and Oroza [66], through a set of experimental workshops informed by the technological disobedience philosophy, show that the process of repair encourages preserving of objects over discarding them.

Repair cultures are a widespread example of making with material constraints. The repair culture of mobile phone technicians in countries such as Namibia, Bangladesh and Uganda are unique in their view of repair skills [33,34,38,39]. In rural areas of Namibia, repair markets emerge within existing marketplaces, providing low-cost mobile phone repairs. The informal repair work of these markets exemplifies a '*tinkering culture*', where reusing components source from broken mobile devices is widespread. In the repair world of Kampala, Uganda, mobile phone repair as seen as a form of technical mastery [34]. Independent technicians, who are unconstrained by repair authorisation, aim to complete a repair by using informal and improvisational techniques. In the repair shops of Dhaka, Bangladesh, repair work is seen as a form of craft and creative repurposing, using existing parts in unexpected ways to complete repairs [1]. Being trained as a technician interests people due to the potential to earn a livelihood [34]. The repair markets inhabit communities with low income and varying educational attainment. This is reflected in their modes of repair. Technical skills are frequently learnt through apprenticeships in more informal repair communities, highlighting the values of peer-learning as informal education. Reusing existing mobile phone parts for repair drives down repair costs, while also allowing for engagement with more skills.

In a more formal making environment, it has been found that material constraints have an impact on the design process [17]. A community woodworking school in the US state of Washington, has had to adapt its making techniques to account for the scarcity of old growth timber [15]. Younger timbers have varying quality; however, its increased availability has made the use of younger timber commonplace with these limitations. Offcuts and warped pieces were salvaged for use in smaller parts of structures. In a different example, students in a university makerspace were tasked with trying to minimise waste products. As such, they were constrained to design a use for these discarded materials as a form of *salvage fabrication* [17]. Initial ideas were soldering and rewrapping offcuts of wiring and reusing milling; and 3D printing offcuts were found to be more of a statement of the history of the making process, rather than items desired for the functional uses. Altering the 3D printing filament to be a reparative glue, using cardboard as a growing medium, and making adhesive from polystyrene packing foam [16]. These have the potential to be utilized as tools within the makerspace. Both these examples illustrate how resource constraints significantly shaped the design process and resulting output; very much in line with how de Bono [6] argues that constraints act as creativity conduits, instead of being an inhibitor or restrictive force.

The practice of making tools in makerspaces out of existing components and materials is a common practice, and a means to address equipment limitations [2]. The ad-hoc making replicates existing, yet unavailable tools—making the result a fluid combination of material and tool. We employed e-waste as our material; given its *material* (i.e., fidelity, constraints) and *relational considerations* [41] (i.e. abundance, value, familiarity etc.) enabled it to be used in a variety of ways. This served as both the material components, and the provocations as part of our generative toolkits [70]; which are central to scaffolding participation and generative design activities.

## 2.3 Making and Participatory Design in Under-resourced Communities

Making, alongside telling and enactment, is identified as one of three central characteristics of participatory design tools and techniques [7,68]. Sanders and Stappers [70] stress on the importance of making within this triumvirate; arguing that "*we really cannot separate making from telling and enacting. We have seen in practice that people make artifacts and then readily share their stories about what they made, or they naturally demonstrate how they would use the artefact (if it is intended to be a representation of something concrete.) Taken in isolation, the artefact may say very little or remain highly ambiguous. In fact, this ambiguity is intentional, as it generates opportunities for creativity, expression, and discussion. The meaning of the artefact is revealed through the stories told about it and the scenes in which it plays a role.*" This stresses the importance that making as an activity has within a participatory setting. *Making* also enables externalisations of concepts; seen as a fundamental feature of all design [11,22,42]; allowing for multiple stakeholders to engage with the same material constructs, where what is made serves as a boundary object [72].

The maker movement has been touted to bring educational, innovative, entrepreneurial and wellbeing oriented benefits [12,35,44,48,52,83]. With its existing outreach (Etsy, Quirky, Thingiverse), funding options (Kickstarter), and technology platforms (Arduino, Raspberry Pi, 3D printers), the maker movement has established an economic model that is worth billions of dollars. There have been recent calls to broaden the scope of making by involving marginalized populations [82]. Recent HCI studies have taken into account the making practices of women [25,30], older adults [65,75,84,88], refugees [73], low SES community members [78,87], among others. A set of studies conducted in an Australian men's shed involving older retired men [88] and an e-waste recycling centre involving unemployed individuals [87] have shown that people's involvement in making activities and social setups within which making takes place have mental health, self-efficacy and societal level benefits. Harrington et al. [28,29] have cautioned that when engaging with low-income communities, researchers should consider the history of research sites as well as evaluate the design outcomes on localized community matrix rather than having a 'corporate' focus. Studies like these have encouraged our current work to engage in participatory design activities within makerspace-like setups.

Frauenberger et al. [26] conducted a series of participatory design sessions with autistic children to co-design smart things for use in their everyday life. As each session worked with a single child, the participatory design methodology was adjusted to suit the child. The focus was more on ideation and the initial design stages, with the items then being separately realised as a prototype. Thus, although the children engaged in the design process of making, they did not explicitly become makers in a physical sense.

A DIY workshop series was conducted with people with disabilities, aiming to explore the empowering potential of making through participatory design [55]. In the series of workshops, the first three taught introductory technical skills including 3D printing, laser cutting and electronics, with the following 2 reserved for developing an individual maker project. They also found that the makerspace used was not very accessible. Of the three ideas prototyped, two had explicit links to accessibility in the form of a boccia ball holder for an electric wheelchair, and temperature sensors [15]. This indicates the intersection of assistive technology with technology for personal use.

# 3  Study Design

## 3.1 Study Site – e-Waste Recycling Center

The study was conducted in collaboration with an e-waste recycling centre, that as a social enterprise engaged local community volunteers and WFTD participants to collect, disassemble, and process e-waste. Housed in a large shed (Figure 1) with work benches, the centre collected e-waste from local areas, and accepted people dropping off their old electronics (e.g., PCs, printers, laptops, home appliances, and other domestic e-waste). The volunteers and staff at the centre disassembled e-waste, and in rare occasions encouraged WFTD participants to make new products by salvaging the e-waste materials. The centre developed various repurposed products from re-claimed e-waste materials such as: power-banks from recycled laptop batteries, 3D printers from recycled printers and PCs, amplifiers from recycled music systems and army ammunition boxes, and electric bikes and digital road warning signs. These types of making activities happened in separate areas within the shed, but still visible to all. Staff members of the centre would oversee and manage the progress of these projects and at times involve WFTD participants in making activities. The centre fostered a visibly prominent peer-learning culture, aimed to provide job-ready skills for capacity building and professional development for WFTD participants. Managed by a large non-profit organisation, the centre was funded by the government. The centre also generated income by selling refurbished products, and making new products with salvaged, repurposed e-waste materials, such as selling road flooding warning signs for the local council. Limited financial resources were used for staffing and procurement of low-cost electronics for product development (e.g., Arduino kits and electronic sensors).

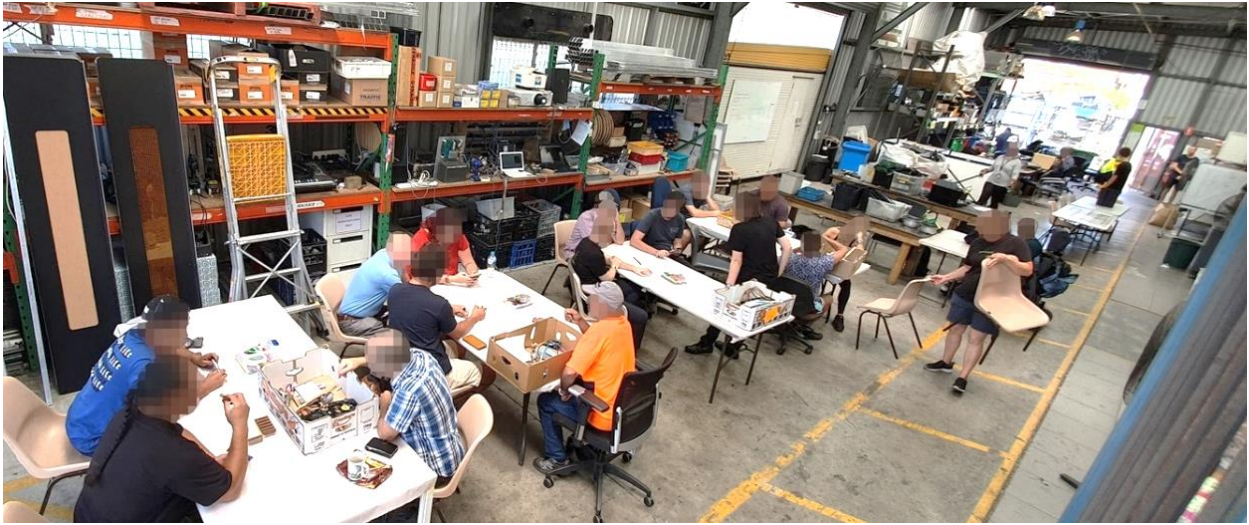

Figure 1. The e-waste recycling centre - study setup (Workshop 1)

## 3.2 Participants

Participants were organised into teams of up to four WFTD participants, and one staff member (see: Figure 2). WFTD participants were skilled at disassembling e-waste and brought unique perspectives from their life experiences, while staff had expertise in repurposing e-waste and making usable products—the overall combination fostering *mutual learning* [62] and reducing *asymmetry of knowledge* [61], both central outcomes of participation in design. A total of 24 participants were engaged over two maker workshops: the first workshop (WS1) had 18 participants (including 4 staff members) and the second workshop (WS2) had 11 participants (including 3 staff members). The participant ages ranged from 18 to 64, however individual backgrounds varied greatly. Before the start of the maker workshops, participants were provided with a consent form (approved by authors' institution) which made issues around the voluntary nature of their participation, audio recording and photo capturing of participants and use of the collected data were made clear to the participants.

The participants were selected using a purposive sampling strategy, (1) ensuring participants belonged to the WFTD cohort; (2) were familiar with the space, materials, and had prior experience disassembling e-waste. The staff volunteered their time to participate and support the WFTD employees in the workshops and had experience of making and repurposing e-waste. Four of the same centre staff members participated in both WS1 and WS2.

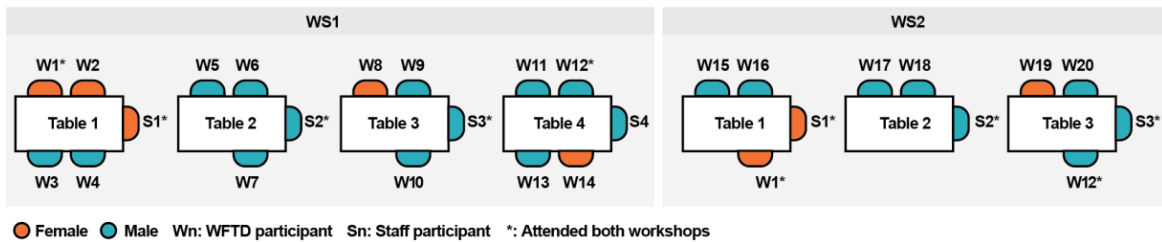

Figure 2. Participants details and table configurations in the two workshops (WS1 & WS2)

It is important to note that all participation was voluntary, to avoid influence of any organisational power imbalances between the centre staff and participants. An active effort was made to create awareness that this was an independent research study—i.e., was not part of the WFTD participant's work, and the centre and staff had equal roles within the ambit of the study. This was supported by a level of trust and rapport built between several WFTD participants and the first author.

## 3.3 The Box: e-Waste Generative Toolkit

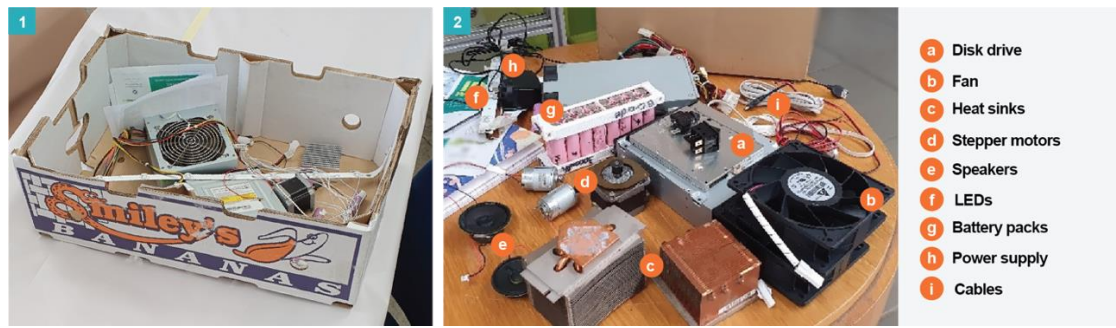

**Figure 3.** (1) 'The Box' – assortment of e-waste materials as presented to participants per table; (2) e-waste materials elements used inside the box; comprising of most commonly available materials at the centre—co-created with centre staff

Participants were provided with an assortment of common e-waste materials (see: Figure 3) as a *generative toolkit* [70] which was prepared prior to the workshop, under the guidance of the centre staff, who had expertise in identifying the most commonly received e-waste materials at the centre. Participants were also advised that they could supplement the toolkit with additional e-waste material from the centre (e.g., Arduinos, printer toner etc.) if required for prototyping. This was to ensure that participants would not limit themselves to specific technologies provided in the box; instead use it as a starting point. As a design material, e-waste is particularly valuable tangible user interface (TUI), because of the embedded *material considerations* [41]—most importantly—constraints (i.e., form, size, function) and fidelity. Constraints have been well established as an effective way to foster creativity (e.g., [32,40]) and lateral thinking—de Bono [6:56] stating "*The constraint is not meant to be restrictive, on the contrary it encourages the effort to find difficult alternatives instead of being easily satisfied.*" The materials also served as *boundary objects* [72] —shared points of reference for participants, scaffolding discussion, and tapping the need to create externalisations from a blank canvas. Sanders & Stappers [70] argue that "*generative toolkits describe a participatory design language that can be used by non-designers (i.e., future users) in the front end of design so that they can imagine and express their own ideas about how they want to live, work and play in the future*" leveraging both visual and verbal literacies [67] through the externalised materials.

## 3.4 Maker Workshop Protocol

The maker workshops were 2-hour co-design sessions where participants designed ideas and developed their first prototypes as a proof of concept. This was followed by a week-long time, where participants were able to develop a functional prototype of their concept. Two of the authors were involved in planning and execution of this study. The overall activity was divided into three stages: (1) *Ideation*; (2) *Making*; and (3) *Build and Reflection*.

Stage 1*: Ideation.* The workshops started off with ideation sessions (30 minutes) which introduced participants the aim of the workshop, materials, and instructions for the activity. Participants were distributed into teams (up to 4 WFTD, and 1 staff participant per table) and, tasked with brainstorming ideas to repurpose e-waste to develop solutions that could address problems within their environment—their household, the centre or broader community. The aim of this phase was to generate multiple design ideas and concepts. Participants were provided (a) the box—with e-waste materials, which served as both resources to work with as well as tangible provocations for ideation; and (b) butchers' paper and markers, to sketch out ideas for the concept. Participants were encouraged to engage with the materials, discuss ideas within their teams and sketch concepts and what they might look like.

Stage 2: Making. Over the course of 90 minutes, participants started working on making their prototypes after the ideation sessions ended. The aim of this stage was to enable teams to create the first version of their proof-of-concept idea. The e-waste toolkit acted as the scaffolding for participation, and participants adapted, modified, and appended to their e-waste toolkits as required while they refined and polished their concepts into higher fidelity prototypes. The role of staff members became important here as they helped in the technical aspects of the prototype building. Following the 90 minutes of co-development session, participants were asked to introduce their ideas and the initial prototype. Teams were able to ask questions to one another about their approaches to build a working prototype.

Stage 3: Build and Reflection. At the end of the two-hour maker workshops, participants were given a week to engage in developing a fully functional prototypes of their ideas. Following the end of the week, the research team conducted a semi-structured interview with one of the participants from each team and discussed their experience of making with e-waste. Participants were asked questions exploring—*What was the underlying motivations behind their concept? What did they learn from the process of making with e-waste? What values were embodied in their concepts? What challenges they foresee in the realisation of some of their ideas? How could their ideas benefit the world around them?* The interviews were conducted at the centre in a communal space; with participants discussing their process in an informal setting on their tables.

## 3.5 Researcher's positionality

The first author had engaged with the centre in various research activities as part of a longitudinal research project, spanning over four years. He worked with the centre and its staff members to plan the overall study and his rapport with some of the WFTD participants enabled a more supporting and creative environment for the study.

## 3.6 Data Analysis

The two-hour maker workshops and follow-up interviews were audio recorded and later transcribed. The participant conversations on each table were independently recorded by multiple audio recording devices. The research team took photos of different teams, their design ideas and final prototypes and captured field notes. Using thematic analysis [8], we started our data analysis by reading through the transcriptions, followed by creating initial coding scheme. We inductively identified codes by iteratively refining codes within our data. This resulted in the development of three themes that summaries our analysis of the data.

## 4 Findings

Our analysis of the study data suggests three main insights into the core values that drove our participants' making practices: Balancing personal vs community needs; Incorporating convenience and productivity; and Re-thinking sustainability and connections. These core values were represented through the 28 design concepts that were generated and discussed by participants during the participatory maker workshops (See Table 1).

Table. 1 Summary of generated concepts & prototypes from both workshops. Bold = Completed prototypes; Focus of ideas: I = individual, C = centre-based and E = community-level.

| Session | Table | Concepts & Prototypes | |
|---|---|---|---|
| **WS1** | 1 | **Screwdriver sharpener** C | System to prevent kids from undoing seatbelts I |
| | | Solar powered sprinklers for community garden E | Fireproof camping tents for fire-ban weekends I |
| | 2 | LED outdoor Christmas decorations I | Automatic gate opener I |
| | | **Sunlight based alarm system** I | Improving blindspot safety mirrors E |

|     |   | e-Waste donation rewards program C | Repurposing and reselling e-waste C |
|     | 3 | Centre cooling system C | Ink/toner sustainable disposal system C |
|     |   | Bins for community e-waste collection C | Repair cafés at centre C |
|     |   | Solar charging hubs for e-scooters E | Centre community promotion C |
|     |   | Phone charging for outdoor events E | |
|     | 4 | **Motorised screwdriver for disassembly** C | Misplaced tools prevention system C |
| WS2 | 1 | **Improving blindspot safety mirrors** E | |
|     | 2 | Dishwasher loading and storage support (for people with disability) I | Real time bus timing displays E |
|     |   | **Visual temperature warnings for rooms, devices and vehicles** I | Bus hailing buttons at bus stops (instead of physical hailing) E |
|     | 3 | Road noise pollution blocking for residences I | Improving community awareness for recycling e-waste C |
|     |   | Disease prevention in farms in local catchment area E | **Representing sound visually** I |

During the ideation process, some groups created a large number of concepts, e.g. Table 3 in WS1 created 11 concepts; while Table 1 in WS2 just wanted to work on a specific idea. Six of these concepts were further developed into functional technology prototypes over the course of the week. For the purpose of this paper we present exemplars that embody unique participant values and exemplify how participants identified and engaged with a diverse range of issues pertaining to individual (I), centre-based (C) and community-level (E) contexts. Our selection of these exemplars was also based on their level of completion, where we found that Table 3 in WS1, for example, had 11 interesting concepts but their final prototype of 'sunlight based alarm system' needed more work. Hence, we did not include it in this paper.

In the following, we present our findings by providing insights into specific prototypes that were designed by participants. For each prototype we present the design process that uses descriptive narrative and data to take the reader through the steps of (i) problem setting; (ii) planning and prototyping; and (iii) making and reflection.

## 4.1 Balancing personal vs community needs

From the beginning of the maker workshops, we had made it clear to the participants that they could think about designing any type of technology that can help their own and community needs. A large number of concepts were designed, which ranged from tacking community-oriented issues around road safety, noise pollution, community gardening and so on to supporting individual needs such as fireproof tents, and technologies for supporting household activities. In order to provide a detailed account of the design process that our participants undertook, we will now discuss an example of a blindspot intersection mirror that was developed by participants who were on Table 2 of the second workshop.

### 4.1.1 Blindspot Intersection Mirrors.

*Problem setting.* Across the workshop session, multiple groups when looking at the *community* context, identified a particularly dangerous intersection near the centre, which had a very problematic blind spot for motorists. W17 and his team at Table 2 (WS2), were discussing the role of adding additional sensors to intersections that had blind-spots to avoid potential accidents. W17 drew on personal experiences of near misses, where he had almost had an accident while crossing through an intersection in his locality. Commenting that overgrown trees at the intersection, created visibility challenges, that the extant convex mirrors were not sufficient to address. W17 was an experienced member of the centre, having dropped out of school, and struggling to find employment, he had been coming to the centre for a couple of years. Aware of the centre's contract with the local council to develop solar flood warning signage from recycled e-waste, he saw this as an opportunity to create a prototype that addressed safety concerns and draw on the centre's maker staff expertise of creating road signs. There was also the very real possibility that if done well, this concept could result in a real-world application—which motivated the group to focus on this idea.

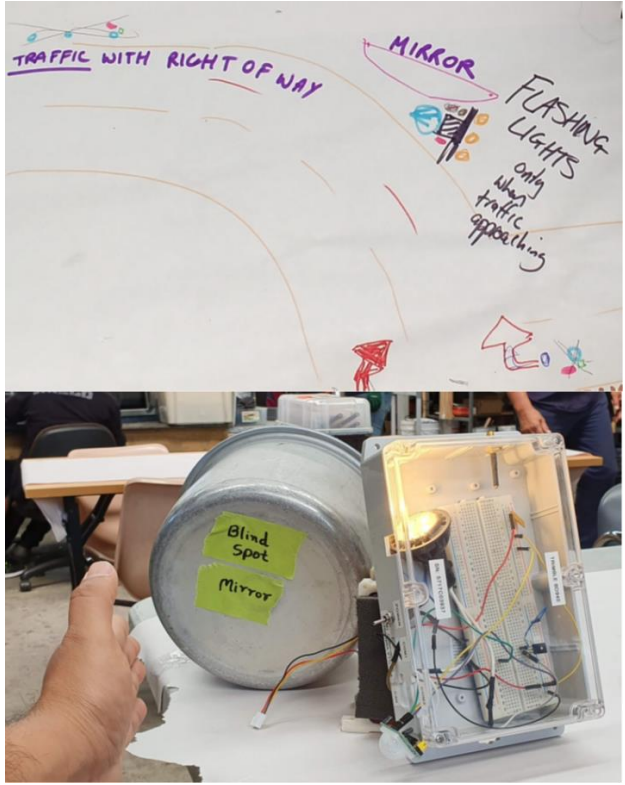

**Figure 4.** Design of blindspot intersection mirror: Sketch of the layout of the blind spot intersection, including mirror placement, traffic information, and positioning of the mirror (top). Participant giving a demo of the sensor-based lighting module (bottom).

*Planning and prototyping.* Having decided on a context, to further flesh out the problematic, the team sketched a diagram of the nearby blind spot, noting traffic flow, road rules and the position of the existing mirror (Figure 4). They reasoned that mirror itself was effective if it was (a) properly maintained—i.e., surrounding foliage was trimmed and mirror kept clean; or (b) larger—covering more surface area of the road in reflections. Hence, instead of redesigning the mirror, the group chose to explore ways to draw attention to the mirror and the dangers of the intersection using a type of sign. W17 said:

> "*I guess it makes it somewhat safer on the road, depending on how bad the convex mirror is. But it's not a new solution, it's just bringing attention to an issue that may be bothering a lot of others… to the solution that is already there*".

It was decided that having flashing lights near the mirror would be the most feasible way to draw attention to it, since lights are already commonly used for traffic signalling. The main challenge was to increase the visibility of the blind spot created. They brainstormed ideas to address this this problem space and reached to a unanimous agreement—that flashing lights would be the easiest indicator to notify both commuters and pedestrians when a car was approaching from the other side. Their idea was that the flashing light would be triggered using a sensor to detect oncoming vehicles and traffic. They were mindful that the sensor should not unnecessarily trigger the light; and concluded that to mitigate this (and chances of errors), light was only needed to be flashed for traffic in one direction, since oncoming vehicles from the other direction would have the right of way. Several sensor ideas were discussed around existing traffic controls, such as using the sensors from traffic lights, or placing motion detectors within the embedded cat's eyes on the road. These were discarded due to the difficulty of actually placing items on the road, and possible wear and lack of durability if the cat's eyes were damaged. As a result, they decided to have a motion detector, attached to the apparatus where the flashing light was. A mockup was made using components from the box and additional materials from the centre, using a directional light that would be activated by a motion-detecting sensor controlled by an Arduino.

*Making & reflection.* In the week following the initial workshop, W17 continued to work on this idea, developing a functional prototype. It consisted of a metallic bowl, representing the convex mirror, and a recycled GPS casing with a transparent cover (Figure 4). This casing was chosen as it is water resistant, thus suitable for outdoor use. The casing held the electronic components: an Arduino, a light and a breadboard connecting components. Attached to the side was a motion sensor and a direct switch to control the lamp. When a hand was placed in proximity to the sensor, the light would turn on (Figure 4). W17 intended to continue iterate and refine the prototype till it is a more effective proof of concept; with the aim to then present it to one of the centre managers as a potential new product to develop.

## 4.2 Incorporating convenience and productivity

This theme in particular highlighted our participants' values around making technologies that can help improve their convenience and productivity around their personal spaces as well as workplaces. Ideas such as screwdriver sharper, motorised screwdrivers, automatic gate opener, amongst others indicated that participants sought to improve their workplace and domestic lives through these types of technologies. In the following, we provide a detailed account of one such prototype – motorised screwdriver.

### 4.2.1  Motorised Screwdriver.

*Problem setting.* Participants at Table 2 (WS1) decided to focus on improving resources that facilitate current practices of the centre. The participants identified two main problems during the *ideation* session (a) the repetitive strain of unscrewing e-waste during disassembly; and (b) tools often going missing. Repetitive unscrewing over extended periods of time, during the disassembly process causes significant strain on hands and arms of workers. It was proposed that some form of motorised electric screwdriver be fashioned, to support workers and cater to the bulk of the initial unscrewing process. Once the screw was loose, manual tools could be used to completely unscrew it. W11 a 38-year-old WFTD participant had joined the centre a few months prior. Although he enjoyed working here, he highlighted a common issue:

> *"Look at these scars and bruises [showing his both hands and fingers]. I guess they will heal, but since I have to use my hands to do this every day, these may not heal that easily."*

W11 then pointed to his hand gloves, saying that he had been using them for protection for some time now; and while the gloves helped initially, they were not a solution. W14 added that while disassembly may sound a simple work, it was very tiring and required a lot of force when getting things done.

Another theme that emerged while identifying problem areas within the centre was lack of female inclusion. There was a common consensus at the table that the centre was not getting enough women to participate in the disassembly process, because it had a visible male dominance. While a lot of women in the past preferred to work in such a laborious place, it was the lack of female membership and community in the space that led existing female members to stop coming to the centre. This discussion led to the emergence of an underlying concern that persisted within the centre, i.e., inadvertent challenges pertaining to inclusivity and diversity within the centre.

Considering both preceding conversations, W11, suggested that one of the ways the centre could make the work easier, simpler, and inclusive for all, was to develop a tool like the *motorised screwdriver*. He reasoned that force is only required to initially loosen the screw, the most strenuous part of the disassembly. Adding, only a small motor would be needed to accomplish this through a screwdriver. And that creating a tool such as this can make disassembly more appealing to a broader audience.

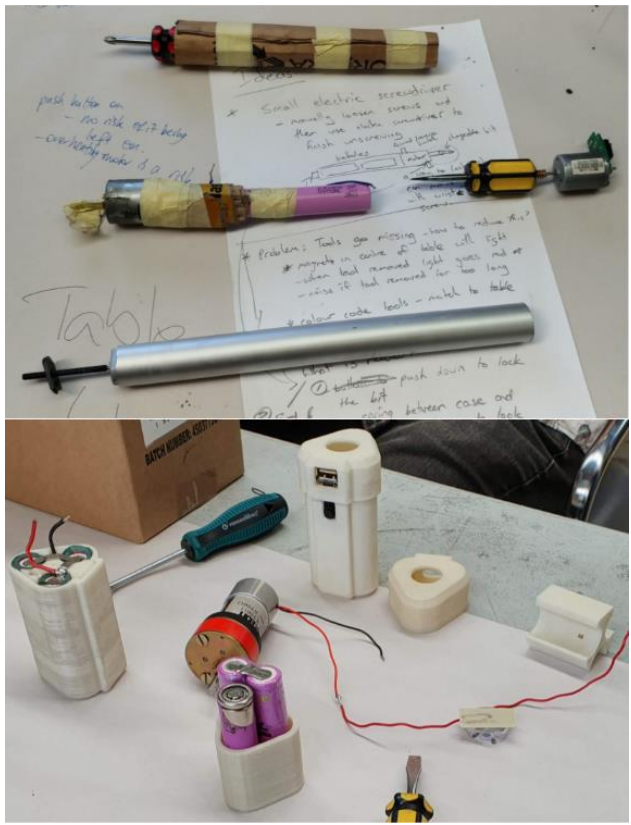

**Figure 5.** Screwdriver prototypes developed within the maker workshop. Top: Ergonomic prototypes (top, bottom), and motorized prototypes (both middle); and Bottom: different housing devices were explored using 3D printed material.

*Planning and prototyping.* The team planned a locking mechanism to allow for the screwdriver to be used manually for initial screw loosening. A push button would be used for on/off to avoid leaving the screwdriver on accidentally and draining the battery. The screwdriver bits were interchangeable, and commonplace and would allow the device to be used for many devices. These ideas were illustrated in a concept sketch (Figure 5). With the more comprehensive concept sketches and availability of the required components in mind, the group chose to continue with this idea over the tool management system. Further prototype concepts were developed based on the core idea of a tube with a screwdriver at one end and batteries housed within the tube. Existing electric screwdriver drills were bulky and difficult to hold up for long periods of time, so the screwdriver design intended to be lighter and more ergonomic. The motors and batteries in the box were a starting point, and from here they collected screwdrivers and other motors from other areas of the centre. Initial prototypes focused on the shape of the screwdriver (Figure 5). By attaching cardboard around the handle of short screwdrivers, participants were able to approximate the size of an electric screwdriver, with batteries being held within the tube. A short piece of aluminium was combined with a screwdriver bit head to make another prototype, allowing for interchangeable screwdriver heads. Motors were also incorporated into some prototypes, first by drilling into a short screwdriver so it could be attached to the shaft of the motor. As S4 – a staff member who was part of this team, had some knowledge of electrical circuits, he connected some spare batteries and a motor to demonstrate the potential movement, forming another prototype. At the end of the session, the group presented a prototype in the intended form factor with a moving motor. By the end of the WS1, the team was able to connect screwdrivers to a motor which can run on recycled batteries. This was quite an accomplishment for the team, who displayed a lot of enthusiasm in creating this functional prototype.

*Making & reflection.* The focus was purely on getting the prototype to work, without much interest in how it looked or felt beyond basic ergonomic/size-based considerations. W11 suggested that the final prototype should be made more presentable and have a brand-new housing with adequate ergonomics to hold it. The team spent the week of the *making* stage of the prototype refining their idea, and creating a more efficient design, with more focus on ergonomics and aesthetics. The team used recycled 3D printers to create a new housing for the screwdriver ([Figure 5](#)). At the follow-up session, the participants presented the how their concepts had evolved. They created a prototype by repurposing an existing 3D printed battery pack case, to which they added a torch attachment. This way they extended their original idea of screwdriver, to have an embedded light source when it was being used; to ensure the screwdriver bit was properly slotted into the screw cavity. They had also found a reversible motor that had enough torque to unscrew tighter screws. The team saw these additions to their original concept as a means to lower barriers of participation, and encourage novice centre workers to engage in the disassembly work.

## 4.3 Re-thinking Sustainability and Connections

The third value that came strongly out of our maker workshops was around how ideas about sustainability need re-thinking and how by incorporating social connections can further support sustainability. A large number of ideas around sustainability and sustainable technologies were discussed in the maker workshops ranging from community-based solar power stations, increasing community awareness on sustainability and waste, repair cafes, to sustainable disposal systems. As the participants were aware of the sustainability ethos of the centre, these ideas came naturally to them. What was interesting here was that participants re-framed sustainability issues as a way to improve social connections within their community. One prominent example here was an art installation that participants on Table 3 developed in the second workshop. We will discuss that example in detail here.

### 4.3.1    Visualizing Sound – An art installation.

*Problem setting.* Participants at Table 3 (WS2) decided to initially explore how might they increase community awareness and engagement by using e-waste materials, however soon pivoted to exploring how to create visual representations of sound. The idea came from W19, who had formally studied fine arts, and was currently pursuing an education in counselling, and wanted to one day do a master's in psychology. W19 was relatively new to the centre, and was not very tech savvy, one of the reasons why she came to the workshop was to learn more technical skills. At the time it appeared to be a very individual-driven motivation—to represent sound as art, with limited discussions as to the rationale for the prototype. However, upon the post-workshop reflection interview (see below) the motivations and thought process was clearly revealed.

*Planning and Prototyping.* Initial conversations around what visualising sound meant, explored both the style of the visual representation, as well as the format of presentation. Below is an excerpt from an exchange of an initial discussion:

> S3:    *"So, how do you think we can visualize sound?"*
> W19: *"Probably getting the sound vibrations into a wiring piece and then onto a computer screen in the form of ECG chart, or something like that."*
> S3:    *"Something similar to a digital visualizer?"*
> W19: *"Ya. But on a canvas."*
> S3:     *"I think we can do that, but what would actually be shown on the canvas? Would it be like lines, or a picture or something?"*
> W19: *"It can be lines, but it can be anything."*
> W20: *"The ones that I have seen had a flat plate with sand on it and speakers at the bottom. It made really cool patterns."*

The group discussed the idea of representing sound graphically, drawing inspiration from electrocardiograms (ECGs), as well as exploring how to transfer graphics onto a piece of canvas. S1, a staff member sitting at Table 1, overheard the discussion on Table 3, and brought over a set of recycled speakers and acrylic beads. S1's idea was to place an acrylic sheet over the speaker, with beads on top of it—when sound is played through the speaker, the vibrations would travel through to sheet to the beads, causing them to move vertically in varying patterns (See: Figure 6). With the aim to explore their initial idea using different material, the group decided to create their first prototype using speakers, amplifiers, music source, acrylic sheets and plastic beads.

Some of this material was not available in the box, so the team members went around the e-waste centre to looking for additional materials. As shown in Figure 6, this particular setup allowed the team to use different music clips and observe vibration patterns on the acrylic sheet. The team also learned that strong vibrations pushed the beads out of the acrylic sheets and the stability of those sheets also needed a bit more attention. In another iteration, the team taped a piece of paper on the acrylic sheet and used printer toner powder on the top the paper. The idea here was to get some form of residual impression on the paper that can be taken home by individuals. The speakers were connected to a phone to produce vibrations from the speakers. It produced some movement of ink, but it was quite messy and did not produce a clear image.

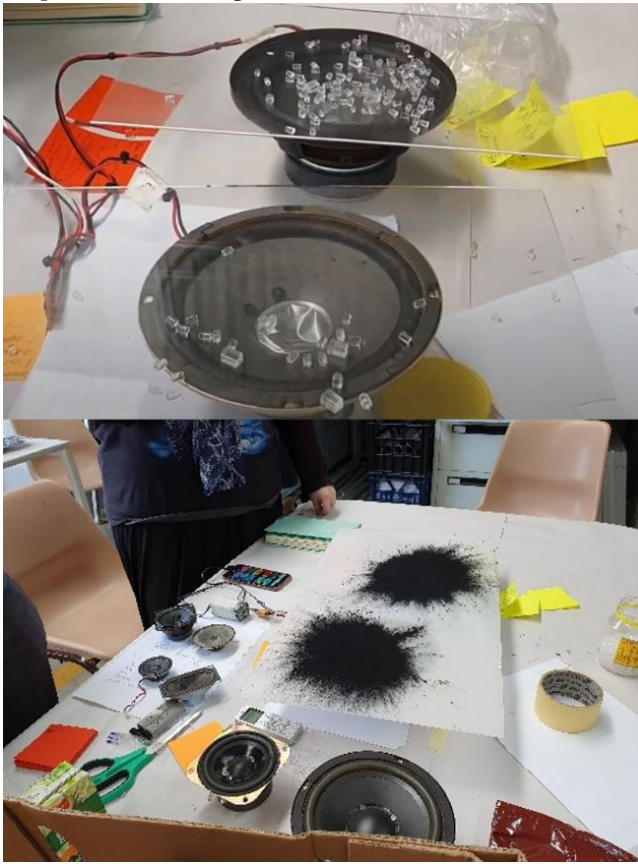

**Figure 6.** Visualizing Sound prototypes: Beads on an acrylic sheet on the top of speakers (top). Printer ink vibrating to the sound coming from the speakers underneath (bottom).

Their next step aimed to better contain the toner. Using an existing acrylic cut-off with holes in it as a frame, butchers' paper was taped to the bottom side. Then, printer toner was poured into the frame, before cling wrap being wrapped around the top to prevent the ink from escaping. The frame was then carefully placed on top of the two speakers, before the music was started. From testing butcher paper and printer paper on the *bottom* of the frame, some patterns emerged, but it was found that having the 'canvas' as the vibrating surface resulted in a lot of smudging, and an unclear artwork. In the next test, the paper was placed on the *top* of the frame, so the toner bounces upward onto it, with cling wrap then wrapped around it to prevent spillage. Different materials of differing thickness were used for the vibrating surface, including cling wrap and butchers' paper. The clearest artwork was created from having the sound vibrate through cling wrap.

While the prototyping process was ongoing team members went back to project's original aims. The discussions concluded that while the current prototype may not solve any specific problems of the community or the e-waste recycling centre, it definitely created a level of engagement among the team members. Several new ideas were discussed by the team where they hinted at moving from printer toner to something less messy, like 3D printing. The mechanics of it were not clear for any of the team members but they showed an enthusiasm for investigating it in the future.

*Making & reflection.* One of the more insightful reflective discussions from the workshop occurred during the follow-up interview for the *Sound Visualisation* project. W19 revealed her motivation for working on this project.

> "*I always wanted to bring sound into a visual representation… It's just that I could never see it, like I could never—I didn't have the tools at that time…*" adding *"most of my artwork was based around motion, tactile feel, vibrations and so on. I research a lot about sound and how sound is in its entirety and abstract concept, like we're just born to just automatically know what certain sounds are…they're not like, you know it's not being taught to us. So yeah, I wanna bring that concept into the visual arts."* She further added that this concept can be further developed as a form of art therapy for individuals who may have mental health problems.

The perceived utility of the prototype—and its application potential for therapeutic purposes was underscored by W19. Being as both a form of inquiry that aims to improve our understanding of sound; but also as a novel means of art-based therapy. Additionally, during this discussion W1, who was listening to the entire conversation and was very intrigued by this idea added a unique application possibility:

> "*I was just thinking… that you can mimic any sort of sound if you wanted to? It'd be unreal to mock a mother's sounds… you know the [toy teddy] bears they got. You can put it into that when babies get taken from their parents—that'd be sick!... because there's nothing they have that can mock a parent's sound*"

This led to the emergence of a conversation about foster children, adoption, and kids being thrown out of their homes living on the streets. W1 also highlighted that there are certain teddy bears that mock a mother's breathing but not commonly available, one which she recently bought for her granddaughter. With the overall aim being to create a sense of comfort for children who for whatever reason are no longer with their parents.

Reflecting on the workshop and toolkit itself, W19 added:

> "*It just helped me visualise it more… and it used to be just a concept in the back of my mind, and I didn't really see it actually being a visualised concept and doing that and like trying to understand y'know parts and all that stuff it actually showed me that it could be possible [to realise]*"

Reaffirming the value of both the structure of the session, as well as the ability to learn about 'parts' and components through the e-waste to realise their idea.

# 5 Discussion

## 5.1 Individual Values & Communal Impact

The three examples presented in the paper highlight how members from under-resourced, low SES communities engaged within making processes, creating technological prototypes that were embedded with their personal values and interests. The generated prototypes embodied the participants' socio-political agenda pertaining to their self-identified problem spaces. This notion of value laden products of making, echoes Sun et al. [75] and Delgado and Callen's [14] findings, where DIY and making were demonstrative of strong political statements of the makers. To Delgado and Callen [14], DIY hacks are not just solutions to specific problems but are open-ended prototypes that present alternative ways of making things that deliberately make publicly visible underlying issues—hence serving as vehicles for political statements.

### 5.1.1 Design Decisions & Political Neutrality.

There is no such thing as a neutral position in design [9,74]; every decision within design practice has a political agenda [61]; laden with the values of those that create them, artefacts too have politics [90]. Taking a case such as the Motorized Screwdriver, which is at face value, is a product that may be easily available off the shelf at local hardware stores. However, when we divorce the motorized screwdriver as merely a product that can be purchased, and delve deeper into the roots of why this concept was explored and developed in situ, we uncover a richer picture of decisions and the rationale that went into the concept. The immediate, very practical need for this tool was to reduce the manual labour that was involved in disassembly—which is routinely done at a massive scale at the centre. Exhibiting their calloused hands and bruised arms to the team, W11 showed the very real physical toll that hours of disassembly take due to constant motor movement, force and fine dexterity used in unscrewing. The seemingly simple, and mundane task, when observed in a household/hobbyist context, became more than a deterrent when compounded over hours. Connecting this to Harringron et al.'s [28,29] studies of engaging with low-income African-American communities, this particular design outcome highlights how the history and politics of the local community was taken into account and how the community's localized matrix informed the design of the motorized screwdriver. Upon further introspection within the team, a deeper theme emerged – pertaining to social inclusion. Although on the outset a tool that has no particular bias towards anyone, the notion that a motorised screwdriver could become the vehicle for inclusion; making disassembly a more inviting task, to a broader audience emerged. W11 saw this as an opportunity to create a more inclusive environment, thinking about the centre's motivations to include more women across the e-waste processing cycle – they thought that this could help in balance the gender split across the centre. Further unpacking the environmental impact, notion of value—within an environment that prides itself in being a cause for good in that they help repurpose wastefulness, simply purchasing new equipment appears to be the antithesis of what the centre stands for. Similarly, the role of value is important to consider, because an off-the-shelf motorised screwdriver might appear as a luxury, a nice-to-have, while the opportunity cost for purchasing it brand new might be too great. A simple, and not even novel concept at face value embodies so much rich thought. The material, appreciation of the difficulties associated with using the materials, the context within which the participants decided to employ it, all factor into its design. We see the statements the artefact makes, in terms of being a vehicle for inclusion, a means for occupation health and safety protection, a means to be sustainable in consumption.

### 5.1.2 Personal Agenda & Communal Impact.

Contemporary design pedagogy often deems designing for oneself, especially within the context of user-centred design, a sin. The *genius design* approach that relies predominantly on the experience and creativity of the designer themselves [58:322], can be seen as presenting the problematic through a myopic, omnidirectional viewpoint. However, when we translate this very approach i.e., designing for oneself, onto a context such as this, where non-designers are drawing upon personal experiences to create changes in the world they'd like to see, we see that this 'design-for-self' is not just beneficial, but also creates ownership, and interest within the process. Moving closer to what Sanders and Stappers [69] highlight as the need for fostering a 'participatory mindset' amongst participants; purchase and interests in the outcome of the design activity. Beck [3] critiqued modern participatory design for moving away from its political origins—this exemplar presents a return to the very deliberate and political stance the participant is taking. There is an acute sense of awareness of the pre-existing relations between the local council and the centre management; and unlike waiting on letters and complaints that are often sent to town planners and council people, W17 is taking this onboard as a way to actively change their lived environment—being the change they want to see in the world around them, by consciously approaching the problematic, cognisant of the possible practical implications of their work. While one could argue the genesis of the idea stems from a personal experience, the widespread implications of it are to help improve the communal experience. The commitment at the end of the session suggesting they will continue to champion this project demonstrates how activities such as this, which merely seed ideas, can actually help foster long-term motivations for participants. We also see that although this is not seen as a novel problem, nor a novel solution—participants flesh out the precise context they want to address, localising their ideas and considerations to the situated environment they wish to change. Considerations such as direction of traffic, wear, and use of familiar elements both signage, and flashing lights demonstrate how participants unpack the problematic, noticing patterns within the space, externalising thoughts onto paper—all demonstrative of very designerly actions that are self-directed.

The idea of visual representation of sound with therapeutic applications, was a result of intrinsic motivation; stemming from a longstanding unrealised goal of W19; something they had always wanted to make (as part of their art); but did not have the technical prowess nor tools to create. Synaesthesia, the merging of different senses, has been of longstanding interest to research (e.g., [49,54])—one of the most curious merging of senses is that of auditory and visual senses—essentially the ability to see sound. Although this concept was the product of the participant's own personal agenda—upon deeper inquiry their underlying motivations surfaced, revealing how they perceived shared communal value within the idea. First, developing a better understanding of sound: sound was seen an abstract concept—something which humans have a pre-configured understanding of at birth, yet we know very little about. Second, as a therapeutic artform, they see tangible sound visualisations as a mechanism for providing therapy for mental health patients. Third, a means of connection at a visceral level—a conversation between W1 and W19 resulted in the discussion of furthering application possibilities if we could mimic sounds from the real world. Recounting a story about how baby bears struggle to adjust away from their mothers, and the role the mother bear's breathing plays, W1 suggested how foster children also face similar challenges in adjusting during adoption. The participants postulated that capturing breathing rhythm patterns of mothers and visualising it for their children can act as a gateway to provide connection and comfort for children (specifically in the case of foster children). This unique application of sound-visualisation highlights just the depth to which this concept can be thought through and possibly applied. Although coming from a place of individual motivation, the structure of the workshop, and the flexible nature of the discussions allowed for more texture to be added to the concepts beyond what was made, to speculating what might be!

### 5.1.3     Contextual Influence.

The ideal of sustainability was strongly implicit across of the ideas and prototypes generated through our maker workshops. There was a bit of rubbing-off happening during the idea creation, where workshop participants chose to focus on ideas that had a strong sustainability statement (in line with the centre's ethos) and had closer connection to the centre's existing projects and activities. This was expected and echoed in other works; for example, in Meissner et al.'s [55] *DIY Abilities* workshop, accessible technology was explicitly mentioned as a potential field to explore which intersected with the participants own experiences around disability. Ideas specifically around the workflow of the centre such as the tool-related products were to be expected, but the discussion of how to promote recycling and increase aware of the centre's work in the wider community was novel. It indicates a strong ethos of sustainability being imbued within the participants, simply from gaining an understanding of these principles from working at the centre. Another concept that was prototypes—the screwdriver sharpener (not discussed in detail in this paper; see: Figure 3) is a more specific example of these values. The aim of the concept was to be able to extend the life of screwdrivers—something valuable to the centre from both a financial and environmental perspective. It was highly unlikely that anyone outside of the e-waste recycling setup would be interested in realizing such technologies, given using a screw driver to the point where its head gets blunt is not a common occurrence in domestic uses of screwdrivers. With throw-away culture, it is easier to just buy a new cheap screwdriver than invest the time and money in repairing one. This speaks more broadly to current consumer culture, where products are not built to last, which to some extent is also reflected in the quantity of e-waste to be recycled.

## 5.2 Strategies for Scaffolding Participation

Making prototypes using the e-waste generative toolkit, resulted in outcomes that resonated with Sanders and Stappers [70] arguments for prototypes— in that "*prototypes evoke a focused discussion in a team, because the phenomenon is 'on the table'…*" and participants had to "*confront theories, because instantiating one [into material form] typically forces those involved to consider several overlapping perspectives/theories/frames*" meaning participants not only had to reflexively engage with the material '*back-talk*' [71] of the e-waste toolkit; but also make all concepts into a format that was visible and accessible to all—"*prototypes confront the world, because the theory is not hidden in abstraction.*" [70].

It is debatable if such a toolkit used in a different situation with a different group of people would have similar findings. As we already discussed, the socio-political issues, personal agendas and the overall context within which the study took place have a strong influence on how such toolkits are used and what kinds of prototype technologies can be designed. While, the e-waste toolkit may not be used, as is, in other situations, specific features and underlying principles can be very useful for utilizing such a toolkit approach for future research projects. In particular, two specific features of the toolkit 1) enabling freedom and 'no fear' of breaking e-waste material, and 2) the use of local knowledge in making the toolkit are very important. These two features clearly helped in scaffolding participation in the WFTD group. Without using the exact same e-waste toolkit, HCI researchers can utilise these two features of the toolkit at a different scale to support engaging participation from various publics. This issue will be discussed further in the later parts of this section.

In the following, we present some generic strategies based on our learnings from this study that can be used to enable participation in a variety of communities including participants from under-resourced communities. These strategies also provide a guide for creating generative toolkits. We acknowledge that these strategies would require further scrutiny to use them as guidelines for supporting participation in under-resourced communities.

**Generative toolkits should be provocative, in addition to being problem-solving.** Traversing through the e-waste toolkits enabled participants to both explore problem-setting (i.e., through the materials) as the materials played as much a role as provocations, as they did for problem-solving (i.e., their utility as material components for prototypes). Hence as a generative toolkit [70] fostering creativity and dialogue amongst participants, the Box was able to structure reflexivity; a core feature of designerly practice [71]. The material constraints of the e-waste were central to fostering creativity (as seen in previous literature [6,32,40]) given that participants explored ideas through the tangible materials, whilst having limited technical knowledge and skills of making. If we compare this study to Dew and Rosner's ecological inversion—the design process explicitly explored uses of specific materials as a way to extend their life and reduce waste [16,17]; whereas in our study e-waste materials serve as a conduit to creative exploration.

**Materials in generative toolkits should invite 'freedom to play' and tinkering.** We believe that the *relational considerations* [41], i.e., the perceived disposability and low value of e-waste, coupled with the participants' familiarity with the material itself, played an important role in scaffolding participation and fostering engagement. With materials like e-waste, there is greater opportunity for creative exploration because participants have no fear of 'breaking' things while exploring and developing the prototypes. Logler et al. [53] in their study investigating young novices working on printer disassembly, that there was reluctance among participants to disassemble the material out of fear of breaking components. Similarly, Khan et al. [41] highlight how material value assessment can act as an inhibitor to participation. Within this study, as participants were aware of the space—the recycling centre and its ethos, they knew that the centre had an abundance of e-waste, and so breaking electronic components during their explorations would not be seen as a negative outcome. Most of the participants were involved in the disassembling at the centre their know-how of specific components and their knowledge of the material flow [16,17] enabled them to find appropriate components for their prototypes. For example, while making the sound visualizer prototype, W19's team members were able to make suggestions around using plastic beads and printer cartridge powder (dried ink). They were able to suggest using components available at the recycling centre that were not provided with the original box as a part of the workshops. This is particularly important as the people who were involved in these workshops were not tech-savvy, nor had any formal or informal trainings around working with electronics, unlike the previous studies [2,16,51,52].

**Generative toolkits should incorporate local knowledge and local context.** While adhocism [2] is seen to be central to enabling DIY in maker cultures, working with e-waste meant that participants were able to grab any suitable component that they see in the centre. The example where acrylic beads were replaced by a dried printer toner for visualizing sound demonstrated that the space itself enabled a strong adhocism through which participants explored creative ideas. Taylor et al.'s [80] work on Community Inventor Days discusses "*reframing making in the familiar*", the idea that for novice makers, working in a familiar problem field is empowering. This intricately related to participatory design, where the user not just engages in the design process but is an expert in that process. At the community level, many participants were aware of the everyday problems and struggles the community faced. For the blind spot signal, W17's group drew a precise sketch of the intersection in the community that caused the most issues for car drivers. Their knowledge of the road rules and environmental conditions greatly informed their design and minimized research requirements. For the centre-specific ideas, most people had worked on disassembly at some stage, and were aware of what would be needed to improve that workflow. This was also evident with individual knowledge of the familiar. W19's work with visualizing sound was related to her own experiences as an artist. This informed what the aims were, and what requirements would be needed to allow for consistent use.

**Incorporate values in scaffolding participation.** Irani [37] reads Latour's [47] conceptualisation of design's value as the translation of objects into things—"matters of concern" where politics of how the material is understood, enable it to have heightened meaning. When we take this approach towards all three cases presented, we see how value-laden the otherwise mundane artefacts are. Prototypes such as motorised screwdrivers become vehicles for inclusion, reducing barriers to participate, and commentary on protection and work effort. Similarly, sound visualisations have been extensively explored both as creative expression and as multi-media representations; however, when paired with the individual motivation and drive of a participation who never thought this was achievable based on their technical prowess — the accomplishment of this outcome becomes more than the outcome itself. The translation of objects into things — when material instantiations of ideas, become matters of concern, of value — we then see how the workshop is successful in fostering a participatory mindset. Truly allowing participants to engage in the production of design.

**Create an ongoing maker culture through participation.** Irani [37] argues that structures like hackathons, "*rehearse entrepreneurial citizenSHIP*" that foster social change; citing structures like hackathons as "*one emblematic site of social practice where techniques from information technology production become ways of remaking culture*" meaning that hackathon-like structures are not just a means to create new ideas and materialise concepts, but also a means to create community and engagement and social reform. Although there is an inherent 'manufactured urgency' [37] created within these social structures, which may not result in concrete tangible outputs—for instance W17 was not happy with the fidelity of their prototype and wanted to continue to work on it beyond the workshop to make it 'presentable' to an external audience; or barring the sound visualisation, the other two ideas are not particularly novel (in that there may be alternatives available already); however the social impact of participation within such engagements has a more pronounced effect. For instance, W19 stating that their prototype "*used to be just a concept in the back of [their] mind*" and "*[they] didn't really see it actually being a visualised concept*" highlights just this — that there is greater social return and accomplishment that is achieved through temporary participatory structures such as this workshop. Being able to understand the 'parts' of the e-waste demonstrates how participants negotiated with the different material assortment and were able to instantiate their ideas.

# 6 CONCLUSION & LIMITATIONS

This work has focused on exploring how to engage under-resourced communities in DIY and making activities and enable them to voice things that are important to them through a set of maker workshops using an e-waste toolkit. In that process, we also unpack strategies for scaffolding participation and engagement of under-resourced communities in *making* using an e-waste generative toolkit. Although we are extending upon existing work relating to making with marginalized groups [55,87] and material constraints [1,16,17,34], our findings uniquely address the combination of these areas and their impact on the design process. This work breaks down established expectations of making culture of one with abundance, and the resultant gatekeeping to making communities. We hope our presentation of a unique perspective on making by engaging under-resourced communities, sheds light on how to scaffold participation using materials such as e-waste, in similar initiatives to develop new technology ideas, sustainable practices, and empowerment for those that engage within the design process. We believe that our work presents new opportunities around exploring different ways and modalities to support participation from marginalised groups. The strategies for scaffolding participation presented in section 5.2 can be applied in feminist makerspaces, for example, where participants engage with different types of fabrics. How would a toolkit for such an environment would look like? How can we design participatory scaffolding for people in different communities from these strategies? If we use the same e-waste toolkit, can we use it in any other way or at a more advanced stage of product development? These are some of the questions HCI researchers can endeavour to explore.

We also acknowledge a limitation of this work in that while the study gave a much-needed voice to an under-resourced community, it did not bring any major changes to their current lives. More work is required to provide a sustainable, longer-term support and facilitation for ongoing making, where knowledge brokers and investors are invited to help participants productize and commercialize prototypes.

## ACKNOWLEDGEMENT

We thank all our study participants, Leah Cowell and other staff members of the e-waste recycling organization for their help and support in our study. Anonymous reviewers have greatly helped in shaping this paper – thank you. This research is funded by the Australian Research Council's DECRA grant DE180100687.